# First principles second harmonic generation of transition metal dichalcogenides and boron nitride alloys: from monolayers and nanotubes to Haeckelites and Schwarzites

Michael C. Lucking, Kory Beach, and Humberto Terrones*

*Department of Physics, Applied Physics, and Astronomy, Rensselaer Polytechnic Institute, Troy, New York 12180, United States*

*corresponding author email: terroh@rpi.edu

**Abstract**

In order to shed light on the second harmonic generation (SHG) of new 2-D systems, first principles methods are used to calculate the second order susceptibility $\chi^{(2)}$ for different types of layered alloys such as monolayers of transition metal dichalcogenide (TMD) alloys, TMD Haeckelite alloys, nanotubes of TMD alloys, hexagonal boron nitride (h-BN) systems which include $B_xN_yC_z$ alloys, BN and $BNC_2$ nanotubes, $B_xN_xC_y$ Haeckelites and BN Schwarzites (porous BN). It is found that the tungsten based alloys possess higher $\chi^{(2)}$ than Mo based at high photon energies, but at low energies, one type of MoSSe dominates. The hypothetical TMD Haeckelites NbSSe and $Nb_{0.5}Ta_{0.5}S_2$ reveal the highest $\chi^{(2)}$ of all the calculated structures. Zigzag TMD alloy nanotubes show higher $\chi^{(2)}$ as the diameter is reduced and approximate to the monolayer for big diameters. BNC alloys exhibit a higher $\chi^{(2)}$ than the h-BN monolayer and are comparable to TMD alloys, except for one case which doubles its intensity. The BN tubes show an increase of $\chi^{(2)}$ as the diameter decreases, similarly to the TMD nanotubes. $B_xN_xC_y$ Haeckelites possess a very high $\chi^{(2)}$ and may shed light on the role of extended defects in nonlinear optical properties. One of the BN Schwarzites exhibits a higher $\chi^{(2)}$ than already known 3-D materials.

**Introduction**

In the last five years, layered materials beyond graphene have attracted researchers due to the exceptional properties of their monolayer systems. The most studied have been the semiconducting transition metal dichalcogenides (STMDs) such as $MoS_2$, $WS_2$, $MoSe_2$, and $WSe_2$, whose monolayers possess a direct band gap and strong photoluminescence (PL) [1-6], valley polarization [7-14], strong excitonic effects [15-23], and strong second harmonic generation (SHG) [24-33]. The case of SHG in hexagonal boron nitride (h-BN) has also been studied, but less intensively, perhaps due to the large direct band gap of around 6 eV [24,34].

Second harmonic generation (SHG) is a nonlinear optical process in which a material interacts with an incident electric field in such a way that the frequency of incoming photons is doubled by the presence of the material [35-37]. SHG is a one of several different nonlinear optical processes that can occur in materials subject to high-energy irradiation; these include sum-frequency generation (SFG), difference-frequency generation (DFG), and optical rectification [35,36]. All of these processes emerge from higher order expansion terms of the polarization density in which the optical susceptibility tensor $\chi^{(n)}$ corresponds to the nth order set of nonlinear optical processes. Second harmonic behavior in a material is dictated by the intensity of $\chi^{(2)}(2\omega,\omega,\omega)$, a 27-component tensor, where the incoming photons $\omega$ are doubled to $2\omega$. Many components of $\chi^{(2)}$ usually vanish due to symmetry considerations for a given material; moreover, a potentially useful property of SHG is that any material with inversion symmetry will have identically zero second harmonic intensity [35].

Although the second harmonic properties of pristine STMDs have been well studied, the effect of having curvature or alloy phases on the SHG requires better insight. Le et al., have been able to synthesize $MoS_{2(1-x)}Se_{2x}$ alloys by chemical vapor deposition (CVD), finding that the SHG is more efficient in these systems[33]. In the present account we address the role of alloying monolayers by examining the SHG through first principles methods which allow the calculation of the second order susceptibility $\chi^{(2)}$. Although we do not include excitonic effects in our calculations, which play an important role in the intensities of $\chi^{(2)}$ [38], we can provide a good approximation on how the second order susceptibility behaves in alloyed TMD materials mainly due to the fact that DFT-LDA exhibits a band gap very close to the optical band gap. In the case of h-BN we have shifted our conduction band to the experimental optical band gap, and for the BN Schwarzites we have shifted the bands in the same proportion as the shift in h-BN. The systems we have considered are the following: different TMD monolayers including chalcogen alloys (MoSSe, WSSe), transition metal (TM) alloys ($Mo_{0.5}W_{0.5}S_2$, $Mo_{0.5}W_{0.5}Se_2$) and 8-4 Haeckelites with Nb and Ta (NbSSe and $Nb_{0.5}Ta_{0.5}S_2$), nanotube alloys (out of TMDs and h-BN), BN Haeckelites [39,40] and BN Schwarzites which are porous BN 3-D crystals with negative Gaussian curvature [41-43]. Though some of these nanostructures have not yet been found experimentally, the results obtained could shed light on the role curvature and alloying play in nonlinear optical properties of layered materials and may motivate experimentalists to synthesize them.

Our results reveal that by alloying TMD and BN layers, the $\chi^{(2)}$ response improves in particular ranges of energy which makes them attractive for robust nonlinear optical devices. Surprisingly, the hypothetical TMD Haeckelites (based on Nb and Ta) of the type 8-4 [41] reveal the highest $\chi^{(2)}$ of all the cases studied here. Also $B_xN_xC_y$ Haekelites of the type 5-7 [39,40] show a very high $\chi^{(2)}$. Alloying BN layers with carbon, besides lowering the band gap, enhances the $\chi^{(2)}$ response. Another interesting result is that BN Schwarzites possess a smaller band gap than h-BN and a higher $\chi^{(2)}$ than any of the 3-D materials found so far [44]. Therefore, the presence of negative Gaussian curvature in BN enhances the nonlinear optical response. This supports the experimental finding that curvature effects can make graphene and bilayer graphene possess SHG signal [45]. Consequently, the different types of Gaussian curvature (positive or negative) play an important role in the nonlinear optical properties of the layered system.

**Methods**

Density Functional calculations are preformed using the LDA functional [46] in the ABINIT code[47,48] with the projector augmented wave (PAW) potential method[49-51]. The PAW potentials for the transition metals Mo and Nb include (4s, 4p, 5s, 4d) electrons in the valence, while the potentials for W and Ta include (5s, 5p, 6s, 5d) electrons in the valence. For the chalcogens S and Se, the PAW potentials include (3s, 3p) and (4s, 4p) electrons in the valence respectively. For the first-row elements B, C, and N, the PAW potentials include (2s, 2p) electrons in the valence. The wave functions are expanded in a plane wave basis up to a cutoff energy of 408 eV. The theoretically determined lattice constants were used for all materials. A $\Gamma$ centered 12 x 12 x1 k-point grid is used for the ground state calculations for the monolayer unit cells of the simple hexagonal lattice, which are 20 Å and 15 Å long in the perpendicular direction for the transition metal and $B_xN_yC_z$ layered materials respectively. Isolated nanotubes were placed in square lattices with more than 15 Å separating the tubes. Seven k-points were used in the periodic tube direction for the calculation of the ground state. The porous P8-0 and G8-0 BN Schwarzites are evaluated with a 4x4x4 and 3x3x3 k-point mesh, respectively. All atomic structures are relaxed until the forces are less than 0.01 eV/Å. For h-BN and the BN Schwarzites, a shift of the conduction band is carried out to match the experimental optical band gap of h-BN. In the case of TMDs this shift is not necessary since LDA provides a good approximation of the optical band gap.

The $\chi^{(2)}$ tensor components [37] were calculated using the ABINIT code. 65 and 100 conduction bands are included for the monolayers and nanomaterials exhibiting curvature respectively. A k-point mesh of 48 x 48 x 1 and 48 x 1 x 1 is used to obtain the wave functions for the optical calculation for the monolayer unit cells and nanotubes respectively. A 6 x 6 x 6 and 5 x 5 x 5 k-point mesh is used for the porous P8-0 and G8-0 BN Schwarzites respectively. A smearing of 0.0544 eV is applied to the optical spectrum to obtain smooth plots. Magnitude of $\chi^{(2)}$ is dependent on the volume of the system, but for 2D systems only the area is well defined causing the calculated intensity to depend linearly on the amount of vacuum included in the supercell. To obtain a value of $\chi^{(2)}$ that does not depend on the amount of vacuum in the supercell, we scale the values by $d_{cc}/l_z$, where $d_{cc}$ is the layer thickness and $l_z$ is the length of the cell in the z direction. Using the experimental cells as a reference, 6.3Å and 3.3Å are chosen for the monolayer thickness for the TMDs and BN structures respectively. Nanotubes are one dimensional, so our scale must be $A_{tube}/A_{cell}$, where $A_{tube}$ is the cross section area of the tube defined by the circle of the outermost chalcogen atoms and $A_{cell}$ is the area of unit the cell perpendicular to the tube axis.

**Results and discussion**

**Second harmonic response in semiconducting TMDs monolayers and alloys**

The lattice constants and band gaps of the pure and alloyed TMDs in this study are shown in table 1. Two types of chalcogen alloys are considered: The first one, labeled MXX(V), which segregates the different chalcogens into different layers of the TMD trilayer structure (Figure 1a). The other alloy, labeled MXX(H), separated different vertical chalcogen pairs in the in-plane direction (Figure 1b). It is worth mentioning that recently the MoSSe(V) has been synthesized [52].The TM alloys (Figure 1c) are constructed in the orthorhombic cell with two $MX_2$ units and have alternating "x" directional chains of the same TM atoms in the "y" direction. The resulting zigzag chains of TM atoms has been observed experimentally in CVD grown alloys [53]. Not surprisingly, the lattice constants of the alloys fall between the two pure materials from which they are formed. The chalcogen alloys possess optical band gaps that fall between the pure materials as expected [52]. However, the metal alloys have band gaps that are slightly lower than either of the materials that form the alloy, indicating band bowing as has been observed previously [54-56]. For some cases, WSSe, WSe$_2$, MoSe$_2$, and MoSSe the conduction band maximum (CBM) moves away from the K point, but the energy difference is small, 0.05, 0.07, 0.02, and 0.002 eV respectively.

|  | $MoS_2$ | $WS_2$ | $MoSe_2$ | $WSe_2$ | MoSSe(V) | WSSe(V) | MoSSe(H) | WSSe(H) | $Mo_{0.5}W_{0.5}S_2$ | $Mo_{0.5}W_{0.5}Se_2$ |
|---|---|---|---|---|---|---|---|---|---|---|
| Lattice Constant (A) | 3.121 | 3.129 | 3.245 | 3.249 | 3.183 | 3.188 | 3.182 | 3.187 | 3.125 | 3.249 |
| DFT Band Gap (eV) | 1.88 | 1.99 | 1.62 | 1.68 | 1.75 | 1.83 | 1.66 | 1.75 | 1.86 | 1.58 |

**Mo and chalcogen alloys**

The calculated second harmonic response with highest intensity $|\chi^{(2)}_{yyy}(2\omega,\omega,\omega)|$ for the $MoS_2$, $MoSe_2$ and MoSSe(V) are shown in figure 2a. For the MoSSe(H) alloy, the upper envelope of all the tensor components is shown in figure 2a; this is due to the fact that the symmetry of the pure trigonal prismatic is broken and other tensor components need to be considered (See supplemental S1), therefore, by showing the envelope the main features of other components can be captured. For the MoSSe(V), the yyy component dominates all the intensities and the envelope function is not necessary (See supplemental S1). The onset in the spectra is approximately at half the optical band gap, as expected, and redshifts as one goes from $S_2$ to $Se_2$ (See figure 2a). In $MoSe_2$, our spectra for low energies agrees with recent experiments with an onset at 0.8 eV and peak at 0.95 eV, which was attributed to excited excitons [57] (See figure 2a and supplemental information S2 with low smearing).There is a second peak at 1.36 eV and 1.3 eV for $MoS_2$ and $MoSe_2$ respectively that is absent in MoSSe(V) (See figure 2a). The peak at approximately 1.6 eV that has been attributed to excitonic resonance[58,59] for $MoSe_2$ appears to be slightly blueshifted in our calculation. This peak is less intense than the one at 1.3eV, but its intensity will likely increase when excitons are included in the calculation. However, MoSSe(V) has the largest peak at 1.76 eV. The Sulfur containing materials all have peaks in the range (1.7 – 1.9 eV), but there is a very small signal from $MoSe_2$. The spectrum for MoSSe(H) exhibits the highest intensity of all Mo-chalcogen alloys at low energies in the range 0.8-1.1 eV and is comparable in intensity with the MoSSe(V) at 2.6 eV, but blue shifted. In order to compare with experimental results, $MoS_xSe_{1-x}$ alloys with 31% Se and 50% Se have been calculated. It is found that as Se increases the $\chi^{(2)}$ intensity also increases as has been demonstrated experimentally[33] (See figure 3).

**W and chalcogen alloys**

The second harmonic response $|\chi^{(2)}_{yyy}(2\omega,\omega,\omega)|$ for the W and chalcogen alloys are shown in Figure 2b. As in Mo chalcogen alloys, the upper envelope of WSSe(H) is shown in the figure to consider all $\chi^{(2)}$ tensor components that may play a role in the optical response (See supplemental S1). Interestingly, there is no onset at half the optical band gap for $WS_2$. All other W based TMDs have a second harmonic response at half the optical gap, though not as large as is seen in the Mo based systems. $WS_2$ has the largest response in the 1.4 – 1.6 eV range, and the peak is redshifted and the intensity is reduced with the incorporation of Se. This is at odds with previous calculations that only included the interband components which predict that $WSe_2$ has a higher response [60]. The WSSe(V) alloy has the same peak position as $WSe_2$ and its $\chi^{(2)}$ does not deviate from that of the latter appreciably at energies below 1.5 eV. At higher energies, the second harmonic response of WSSe(V) is greatly enhanced with respect to $WSe_2$ and is the largest of the W based TMDs. The response of WSSe(H) is very similar to that of MSSe(H) in the sense that at energies in the range 0.8-1.1 eV possesses the highest intensity of all W-chalcogen alloys. One notable difference is the onset of the response at half the band gap which is absent in $WS_2$.

To shed light on the absence of a signal from $WS_2$ at half the optical band gap energy, we do an in-depth comparison with $MoS_2$ in figure 4. The origin of the second harmonic response at half the band gap energy in $MoS_2$ comes from the real part of $\chi^{(2)}$, which is zero for $WS_2$ (see insets in figure 4). The intraband $2\omega$ term is approximately zero for $MoS_2$ at half the band gap, while it has a finite negative value for WS$_2$, which cancels with the positive interband term. The larger intraband term in WS$_2$ signifies that the electrons at the band extrema at K are closer to the limiting case of free electrons than in MoS$_2$ (See figure 4).

**Transition Metal Alloys**

The second harmonic response of the TM alloys are shown in figures 2c and figures 2d. Due to the symmetry breaking and as in the case of the chalcogen alloys, the upper envelope of $\chi^{(2)}$ is shown for Mo$_{0.5}$W$_{0.5}$S$_2$ and Mo$_{0.5}$W$_{0.5}$Se$_2$ (See supplemental information S3 to see all the tensor components). The $\chi^{(2)}$ of Mo$_{0.5}$W$_{0.5}$S$_2$ at half the band gap is approximately equal to that of MoS$_2$, suggesting that the band extrema closely resembles this material. The peak at approximately 1.4 eV coincides with the peak for MoS$_2$, but the intensity is reduced. The next peak at approximately 2 eV is halfway between the peaks for MoS$_2$ and WS$_2$, though the intensity is close to that of MoS$_2$. At energies higher than 2.5 eV, the intensity of $\chi^{(2)}$ for the alloy falls between that of the two pure phases. The intensity of Mo$_{0.5}$W$_{0.5}$Se$_2$ follows the intensity of MoSe$_2$ closely up to photon energies of 1.5 eV. There is a small enhancement and redshift in the onset of the spectrum, likely due to the 0.1 eV reduction in the band gap. The peak at approximately 1.3 eV has a longer tail than MoSe$_2$, but a slightly reduced intensity. The peak at 1.7 eV seems to be an average of the peaks at 1.6 and 1.8 eV for the Mo and W pure phase respectively. Above 2 eV, the intensity is close to that of MoSe$_2$, it is not harmed by the much lower intensity of WSe$_2$.

**TMD Haeckelites alloys**

The hypothetical TMD Haeckelites [41] are also interesting materials. These structures are made of 8 and 4 member rings, but they contain inversion symmetry so the second order nonlinear optical response will be zero, however, alloying the Haeckelites can break the inversion symmetry to potentially give rise to a SHG response. We choose to study the Nb based Haeckelites because they possess a band gap, unlike those made from Mo or W [41]. NbSSe and Nb$_{0.5}$Ta$_{0.5}$S$_2$ alloys were both considered (See figures 1d and

1e). The alloys are constructed such that the atoms alternate in the "x" direction but not in the "y" direction, we have alternation in the "x" direction of the "y" directional chains of similar type atoms. The $Nb_{0.5}Ta_{0.5}S_2$ alloy has a 0.26 eV direct gap while the NbSSe alloy has an indirect gap of 0.46 eV (direct gap 0.48 eV. See figures 5c and 5d). These structures possess $C_{2v}$ symmetry with space group $Pmc2_1$ (number 26). This gives 5 independent nonzero $\chi^{(2)}$ components, zzy=zyz, xxy=xyx, yzz, yxx, and yyy[35] (see figures 5a and 5b). The $Nb_{0.5}Ta_{0.5}S_2$ alloy has a larger response, shown in figure 5b, likely due to the smaller band gap. For both alloys, the zzy component of $\chi^{(2)}$ is smallest, and the yzz is also very small for NbSSe (see figure 5a). The yyy component is largest for NbSSe followed closely by the yxx, while the yxx component is largest for $Nb_{0.5}Ta_{0.5}S_2$: The peak positions are at 0.2 – 0.3 eV for the SHG. NbSSe exhibits a higher response at energies above 0.8 eV. The highest peak for the $Nb_{0.5}Ta_{0.5}S_2$ alloy is nearly 13,000 pm/V, which is over 4x larger than the highest intensity achieved in the traditional trigonal prismatic TMD monolayered structures. Even above 0.8 eV, the NbSSe Hackelite has a SHG response that compares favorably to the highest hexagonal trigonal prismatic TMD materials.

## TMD Nanotubes

We now turn our attention to the second harmonic response of 1D nanostructures, namely nanotubes. Different types of TMD (10,0) zigzag nanotubes were calculated. All armchair tubes have no second harmonic response due to the presence of inversion symmetry. The zigzag tubes have direct band gaps at Γ in agreement with previous works[61,62]. The band gaps are 0.20 eV for $MoS_2$, 0.42 eV for $WS_2$, 0.14 eV for $MoSe_2$, 0.27 eV for $WSe_2$, 0.44 eV for MoSSe, and 0.63 eV for WSSe. The larger band gaps for the chalcogen alloys agrees with previous results[63] and is likely due to the increased stability.

The calculated $\chi^{(2)}_{xxx}(2\omega,\omega,\omega)$ for the tubes are in figures 6a and 6b for Mo and W respectively. The Mo based tubes all have higher and redshifted response compared to the W based counterparts. Both observations are consistent with the smaller gap of the Mo based tubes. The first peak of the TMSSe tubes clearly falls in between the peaks of the pure phases. The intensity is also enhanced with respect to either pure phase, which is in opposition to the general trend of SHG vs band gap found in the flat monolayers. A breakdown of the individual contributions to the $|\chi^{(2)}_{xxx}(2\omega,\omega,\omega)|$ for the MoS$_2$ tube is shown in the inset of figure 6. The large peak at 0.6 eV is mainly due to the imaginary part, from the 2ω intraband component. The interband components become significant at large photon energies, but the 2ω component is cancelled out by the ω components. The intraband 1ω term is small at all photon energies. Compared to the monolayers, the tubes have a redshifted spectrum with a much higher intensity. The intensity of the TMD tubes remains high until 1.4 - 1.7 eV (WS$_2$ - MoSSe), at which point the two have comparable responses (See figure 6).

**Semiconducting BN and BNC$_2$ Monolayers**

Hexagonal BN (h-BN) has a very small second harmonic response (Figure 7), especially when compared to the TMDs. The large band gap is one reason why the response is so small. The DFT-LDA band gap for a h-BN monolayer is 4.61 eV while the experimental optical band gap is around 6 eV [24,34]; this is due to the DFT underestimation of the electronic band gap: In order to compensate for this difference, in our calculations for h-BN and BN Schwarzites we have shifted the conduction bands to match the experimental optical band gap. Interestingly, the band gap and the nonlinear optical properties of h-BN can be tuned by alloying BN with Carbon. In this context, we have chosen B$_x$N$_x$C$_y$ alloys with a reduced band gap [64-66]. In reference [64] three BNC$_2$ motifs are considered, type I, type II and type III. Since type I is a metal, we are not going to consider it in our calculations. Type II BNC$_2$ motif features alternating zigzag chains of Carbon and Boron Nitride (See figure 7) while type III BNC$_2$ exhibits alternating stripes of hexagons that contain 2(B-N) and 2(C-C) units and each hexagon has the same amount of Boron and Nitrogen (See figure 7). Note that type II and type III have been identified in experimental alloys [67]. Our calculated band gap from the type II is 1.62 eV, in agreement with previously published results [64] (See supplemental information S4), however our gap for type III is 1.87 eV, much larger than the 0.5 eV reported previously[64] (See supplemental information S4). A B$_3$N$_3$C$_2$ alloy with a higher band gap of 2.4 eV was also considered (See figure 7).

The upper envelope of the calculated second harmonic response $\chi^{(2)}$ for the BNC alloys is shown in figure 7, all the components are shown in the supplemental information S5. The intensity of the pure hexagonal BN monolayer has an onset around half the experimental optical band gap and is very small in magnitude. In figure 7 we show both h-BN structures, one without the shift to match the optical gap and another with the scissors operator to match the optical gap to provide a more reliable result. Both $BNC_2$ alloys have a giant redshift, as expected from the drastic reduction in the band gap. Moreover, the second harmonic response is greatly enhanced and for the type III and $B_3N_3C_2$ is comparable to the monolayer TMDs. Surprisingly, the type II SHG intensity doubles that of TMDs exhibiting this intensity over a wide range of energies from 0.8 to 3.2 eV. Therefore, all these alloys are good candidates for useful nonlinear optical materials. At lower energies, the second order susceptibility of the $BC_2N$ alloys is significantly higher than that of the monolayer TMDs. At telecom wavelengths, around 1550nm or 0.8 eV, these alloys have an appreciable second harmonic response. The $B_3N_3C_2$ case possesses a higher intensity than type III at higher photon energies with a larger band gap.

**$B_xN_xC_y$ Haekelites**

As seen above in the case of TMDs, Haeckelites have the potential to be extraordinary nonlinear optical structures. The $B_xC_yN_x$ systems allows us to create 5-7 (pentagons and heptagons) structures in addition to the 8-4 (octagons and squares) motifs mentioned above. The structures of the 5-7 $B_3C_2N_3$ and 8-4 $B_2N_2C_4$ along with their calculated second order suceptibility $\chi^{(2)}$ tensor components are shown in figure 8a and 8b respectively. Like the TMD Haeckelites, these materials have exceptionally high second harmonic response. The 5-7 and 8-4 structures have indirect gaps of 1.15 eV and 1.02 eV respectively. The direct gap of the 8-4 Haeckelite is only 2 meV higher than the indirect gap (See supplementary information S4). The smaller gap of the 8-4 is evident in the redshifted $\chi^{(2)}$, which has a peak at approximately 0.5 eV, half the band gap. The first peak in the 5-7 Hackelite is not until 0.8 eV, though it starts to show a significant response at around 0.6 eV, half of its band gap.

**BN and BNC$_2$ Nanotubes**

The $\chi^{(2)}$ of many boron nitride nanotubes have been studied using first principles in such a way that only the direct interband terms were considered[68,69], as well as through tight binding calculations [70]. As shown in figure 9, the SHG intensity tends to decrease as the diameter of the tube increases, which is in agreement with the general trend reported by Guo and Lin [68,69], as well as with our results for TMD nanotubes. Qualitatively, this is a reasonable trend as we would expect the $\chi^{(2)}$ intensity to approach that of the monolayer as the diameter of the tube approaches infinity. However, a closer look at the interband and intraband terms, shown in figure 9 inset, reveals that the intraband terms, which are not considered by Guo and Lin [68,69], dominate in both the real and imaginary parts of the second order susceptibility $\chi^{(2)}$. While the direct LDA band gap of the tubes decreases from 4.3 eV for the (12,0) tube to 3.7 eV for the (8,0) tube, the small redshift in the peak positions does not appear to be a result of this change, as the peaks appear at higher energies than half their respective band gaps. Rather, we must attribute most of the behavior to the complex intraband processes that involve movements along bands that interplay with both the linear response and with interband polarization processes[71].

The nonlinear optical properties of two types of (10,0) BNC$_2$ nanotubes were also calculated. The two tubes that were considered, shown in a side view in figure 10, have the same stoichiometry but different orientations of C-C and B-N bonds with respect to the tube axis. These two nanotube are derived from the two monolayers considered in figure 7, where the type II tube corresponds to a rolled-up type II monolayer and a type III tube corresponds to a type III monolayer. In the type II BNC$_2$ tube the C-C and B-N bonds form zigzag chains around the tube whereas in the type III B-N and C-C bonds are parallel to the tube axis. This difference in bond orientation leads to significant differences in the second harmonic response.

Different components of the nonlinear susceptibility $\chi^{(2)}$, shown in figure 10a, are of interest for the two BNC$_2$ nanotubes. For the type II tube, the yzz component has highest intensity and for the type III tube, the yyy component is strongest. By far the largest peak is the yyy peak for the type III tube at about 1.25 eV (992 nm). Both the type II and type III tubes have significantly smaller direct LDA band gaps (1.54 eV and 1.53 eV respectively) than the BN nanotubes. The $\chi^{(2)}$ intensity of the $BNC_2$ type III nanotube is both significantly higher and significantly redshifted with respect to that of the BN nanotubes; this redshift can at least partially be attributed to the smaller band gap. The yzz component of the type III tube also has a peak near 1.25 eV, albeit much smaller, but it also has two larger peaks at higher energies that must be attributed to more complex features than the band gap transition.

**Porous BN Structures (Schwarzites)**

Hypothetical porous 3-D structures with negative Gaussian curvature, named Schwarzites, first proposed by Mackay and Terrones [72] for carbon materials, have been studied using boron nitride [43]. The negative Gaussain curvature in BN Schwarzites is due to the presence of octagonal rings of alternating Boron and Nitrogen atoms (see figure 11). Following the notation used in reference [73], the G8-0 and P8-0 structures exhibit LDA direct band gaps at Γ of 2.72 eV and 3.16 eV respectively (See supplemental information S6) which have been shifted equivalently to consider the experimental optical band gap of h-BN. These structures possess $T_d$ symmetry which gives one independent $\chi^{(2)}$ term, xyz. The calculated $\chi^{(2)}_{xyz}$ for the porous structures is shown in figure 11. Along with a redshift, due to the decreased band gap of the porous structures, an enhanced SHG intensity is observed. The signal of the G8-0 is higher than the maximum intensity reported for $Li_2CdGeS_4$, one of the highest $\chi^{(2)}$ 3D-materials found so far (See figure 11). The P8-0 structure, shows a lower intensity than the G8-0, but with a nice plateau between 2 and 2.6 eV which could be useful for nonlinear optics applications.

Note that by introducing negative Gaussian curvature via BN octagonal rings, 3-D porous materials with worthy nonlinear optical properties can be generated. As in Carbon, different types and sizes of BN Schwarzites can be calculated, the challenge is thus their synthesis.

**Conclusion**

In conclusion, we have shown that by alloying and introducing curvature in 2-D layered materials new nonlinear optical systems can be obtained. Chalcogen alloys possess high $\chi^{(2)}$ intensities, being the highest the tungsten based alloy WSSe(V) at around 2 eV, however pure MoSe reveals a higher $\chi^{(2)}$ at around 1.3 eV, higher also than $WS_2$ around this photon energy; at lower energies MoSSe(H) dominates. Transition metal alloys of the type MoWS or MoWSe reveal lower $\chi^{(2)}$ than chalcogen alloys, therefore, chalcogen based alloys are better candidates for new nonlinear optical devices. Regarding TMD nanotube alloys, we have demonstrated that Mo based nanotubes possess higher $\chi^{(2)}$ than W based: Both systems show lower band gaps than the flat monolayers and higher $\chi^{(2)}$ at lower photon energies. TMD alloy nanotubes look promising for nonlinear optical devices at low energies. Though $MoS_2$ and $WS_2$ nanotubes were synthesized 25 years ago [74-76], new efforts in the synthesis need to be made, and new integration techniques need to be implemented to use their nonlinear optical properties in new devices. The same applies to BN zigzag and $BNC_2$ nanotubes which exhibit high $\chi^{(2)}$, though in $BNC_2$ nanotubes the $\chi^{(2)}$ is higher than in pure BN zigzag tubes. Our results indicate that flat layers of BNC alloys possess much higher $\chi^{(2)}$ intensities than pure flat h-BN. To synthesize BNC alloys we suggest the strategy to start from already grown chemical vapor deposition BN monolayers and then add the carbon from a carbon source. Starting from graphene has leaded to segregation of BN and Carbon [77] which may not be suitable for nonlinear optical properties. Surprisingly, TMD Haeckelites exhibit the highest $\chi^{(2)}$ of all the systems we have calculated. In this context, among the BN systems studied, BN Haeckelites also reveal a very high $\chi^{(2)}$, not as high as in TMD. We have found that the introduction of defective patches in an ordered way enhances the nonlinear optical response. In fact, the defects in the Haeckelites can be seen as a combination of positive and negative Gaussian curvature patches in the same proportion to balance the curvatures producing a flat layer [39,40]. Experimentally, ion irradiation of TMD and BN alloy systems, at high temperatures, might lead to structures with Haeckelite patches that could greatly increase the $\chi^{(2)}$. It is worth noticing that the 8-4 patches and 5-7 patches have been observed in grain boundaries in

MoS$_2$[78,79]. Our results disclose that negative Gaussian curvature BN Schwarzites (Porous BN) exhibit higher $\chi^{(2)}$ than 3-D known systems. In general, we have shown that higher $\chi^{(2)}$ responses correspond to lower band gap systems, a similar trend is found in 3-D semiconductors [80-82] and non-alloyed TMDs [38,83], but also intraband and interband effects are relevant. In addition, our results reveal that curvature plays a an important role in the $\chi^{(2)}$ response.

**Acknowledgements**

We are grateful to the National Science Foundation (EFRI-1433311). The supercomputer time was provided by the Center for Computational Innovations (CCI) at Rensselaer Polytechnic Institute and the Extreme Science and Engineering Discovery Environment (XSEDE, project TG-DMR17008), which is supported by National Science Foundation grant number ACI-1053575.

**Figures**

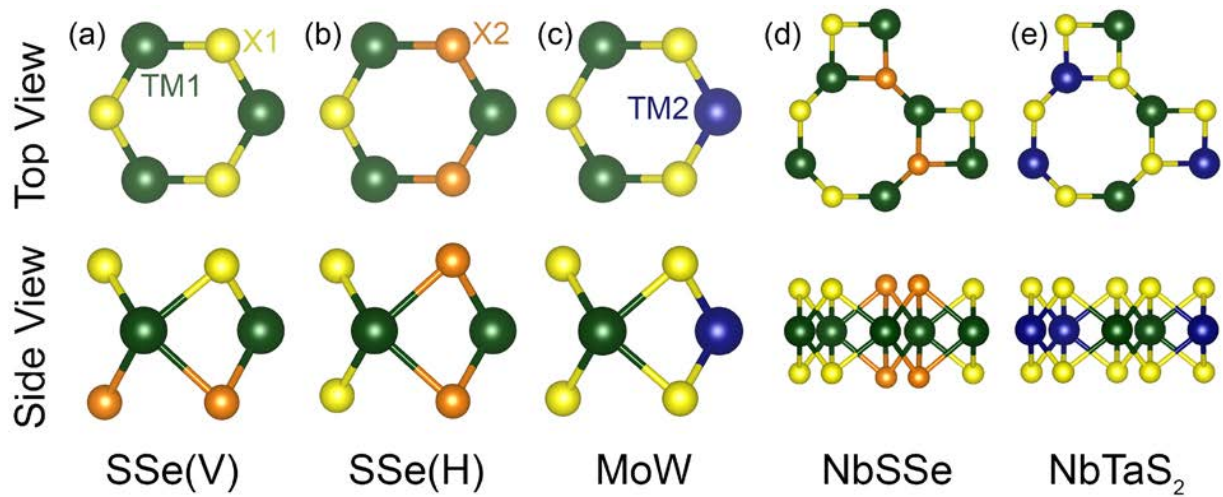

Figure 1. a) and b ) Chalcogen alloys SSe(V) and SSe(H). c) TM alloy. Haeckelite 8-4 alloys d) NbSSe and e) NbTaS$_2$.

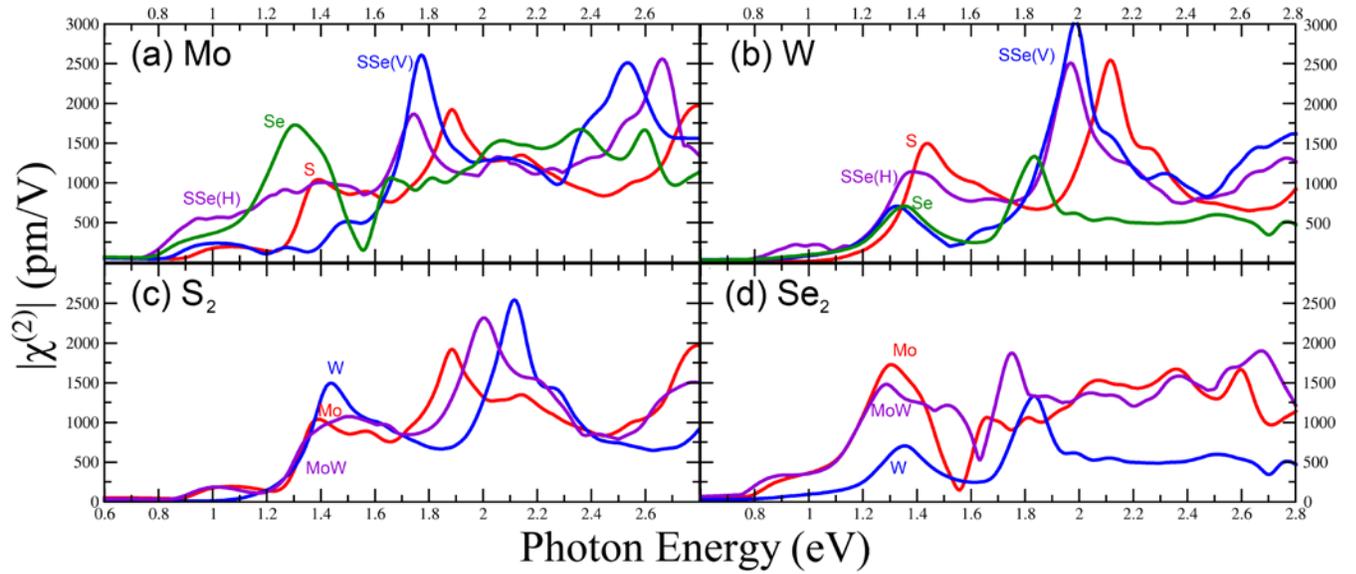

Figure 2. Calculated second Harmonic response of the monolayer hexagonal TMDs. Top figures show the effect of changing the chalcogen for (a) Mo materials and (b) W materials. For $MoS_2$, $MoSe_2$ and MoSSe(V) the $|\chi^{(2)}_{yyy}(2\omega,\omega,\omega)|$ is shown, and for the MoSSe(H) the upper envelope of all components of the $\chi^{(2)}$ tensor are depicted since not only the yyy component dominates at low energies. Similar case happens with WSSe(H) in figure (b). Bottom Figures show the effect of changing the TM for (c) S materials and (d) Se materials.

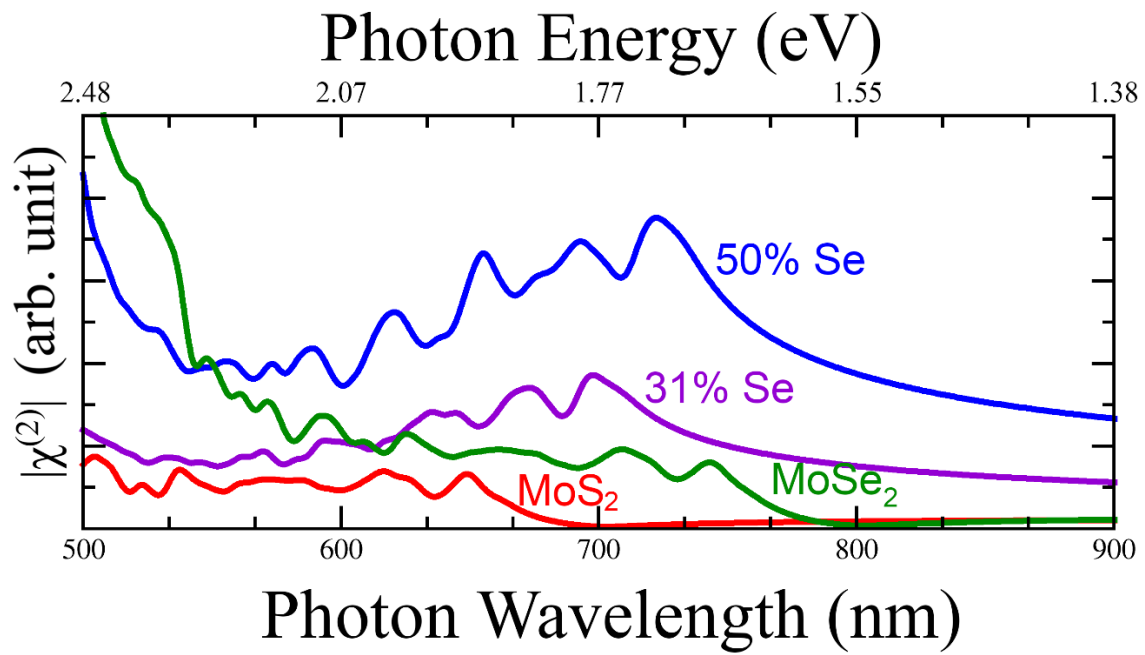

Figure 3. SHG of random $MoS_xSe_{1-x}$ alloys are compared with the two pure phases. An enhanced signal at 50% composition is in agreement with experiment [33]. The energy and wavelength correspond to the frequency doubled photon.

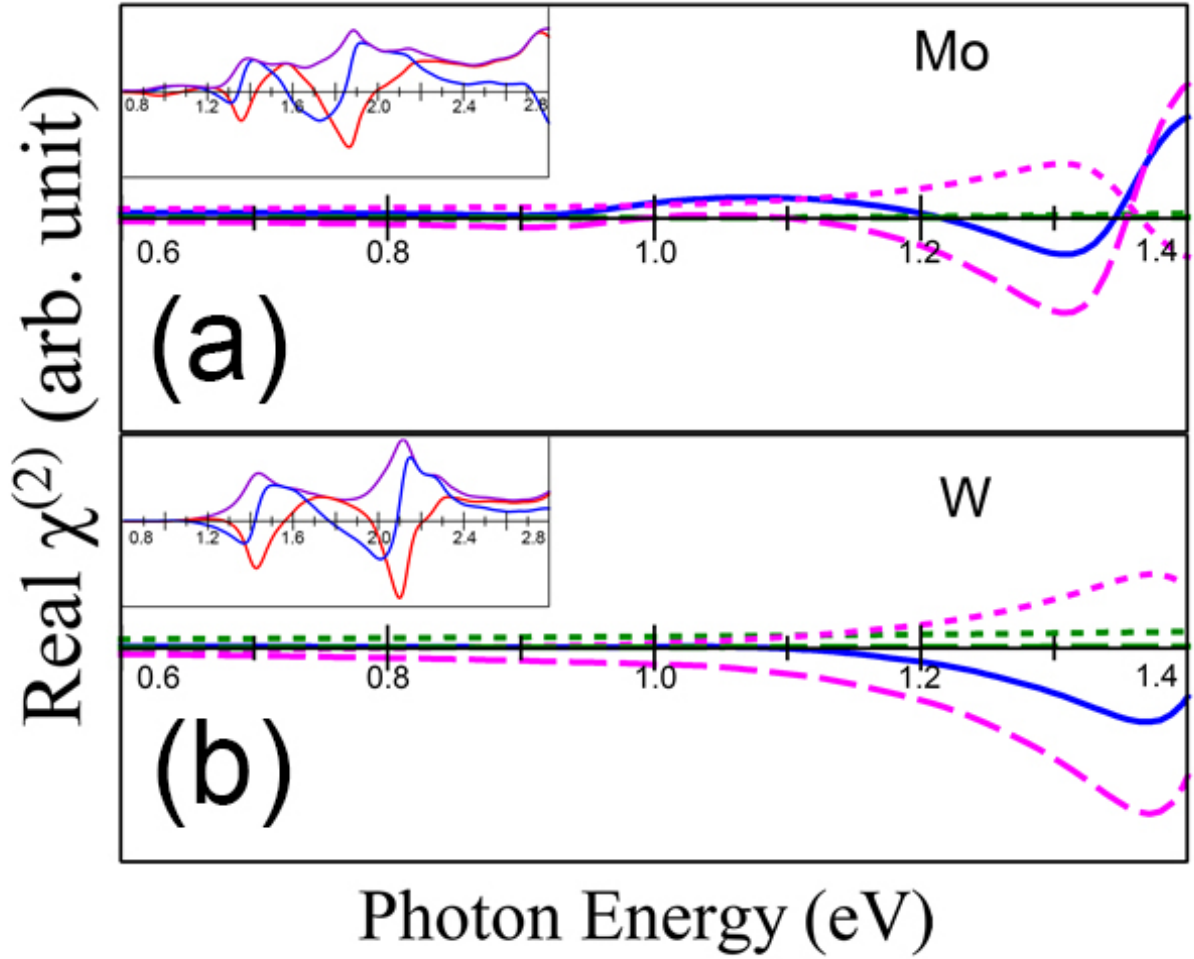

Figure 4. Real components of $\chi^{(2)}$ for (a) MoS$_2$ and (b) WS$_2$. Dotted lines are the interband terms and the dashed lines are the intraband terms. Green lines show the 1ω component and pink lines show the 2ω component. Inset: Comparison of $|\chi^{(2)}|$ (purple) with its real (blue) and imaginary (red) parts.

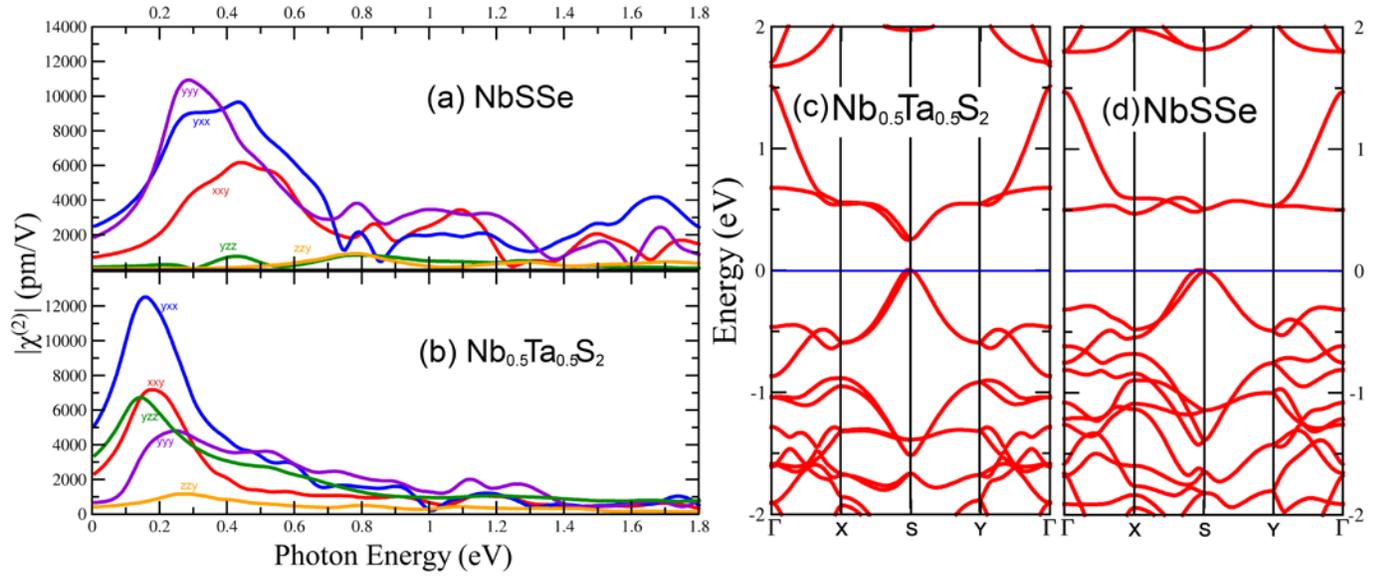

Figure 5. Large Second harmonic response for TMD Haeckelites (a) NbSSe and (b) $Nb_{0.5}Ta_{0.5}S_2$. (c) Band structure of $Nb_{0.5}Ta_{0.5}S_2$ Haeckelite. (d) Band structure of NbSSe Haeckelite.

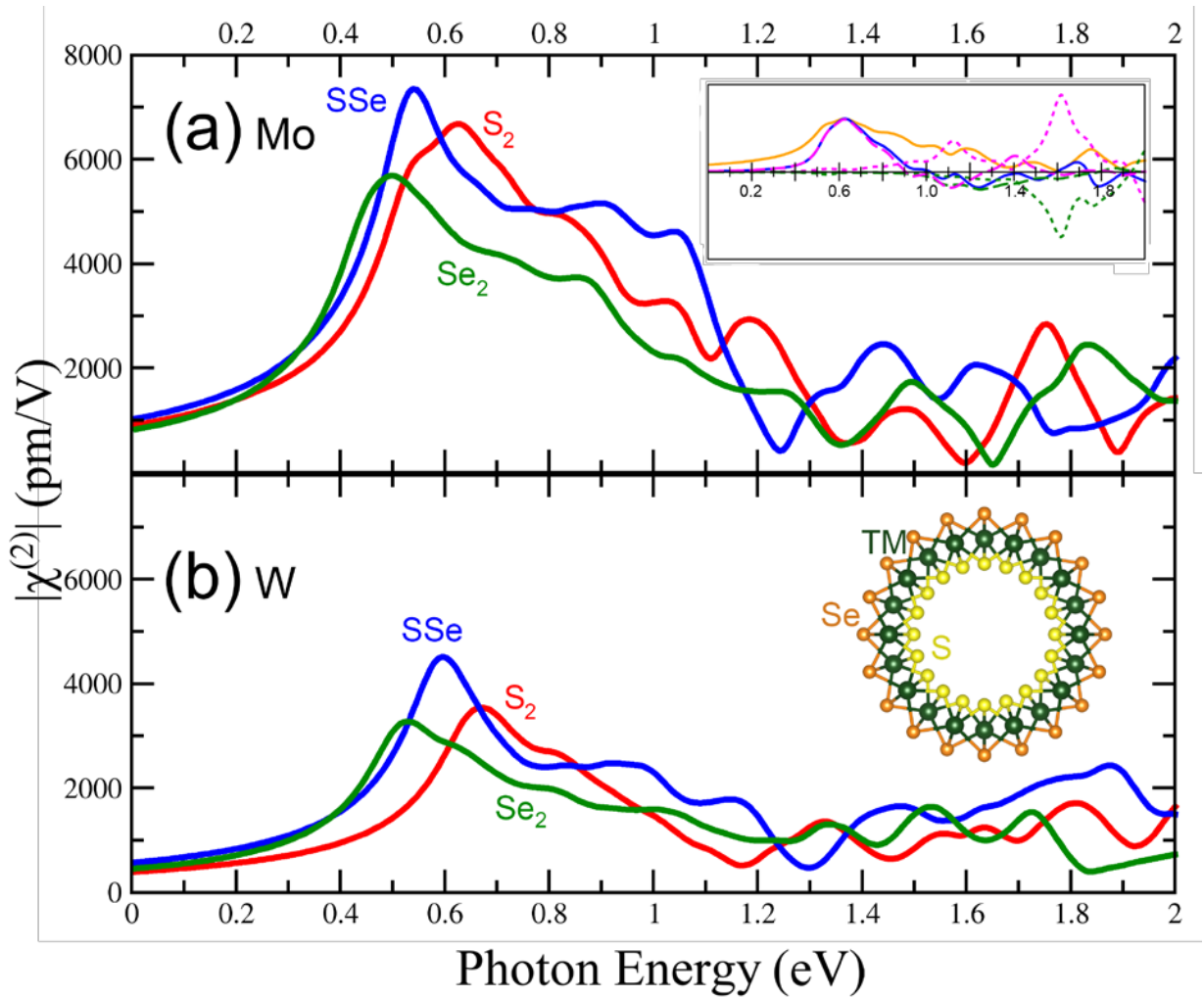

Figure 6. $|\chi^{(2)}_{xxx}(2\omega,\omega,\omega)|$ of (a) Mo TMDs tubes and (b) W TMDs tubes. Tubes with $S_2$, SSe, and $Se_2$ chalcogens are shown in red, blue, and green respectively. Inset: Comparison of the imaginary components of $|\chi^{(2)}|$ (purple) for the $MoS_2$ tube. The total imaginary part is shown in blue. Intraband terms are in green and Interband terms are in red. Dotted lines represent $1\omega$ terms and dashed lines represent $2\omega$ terms.

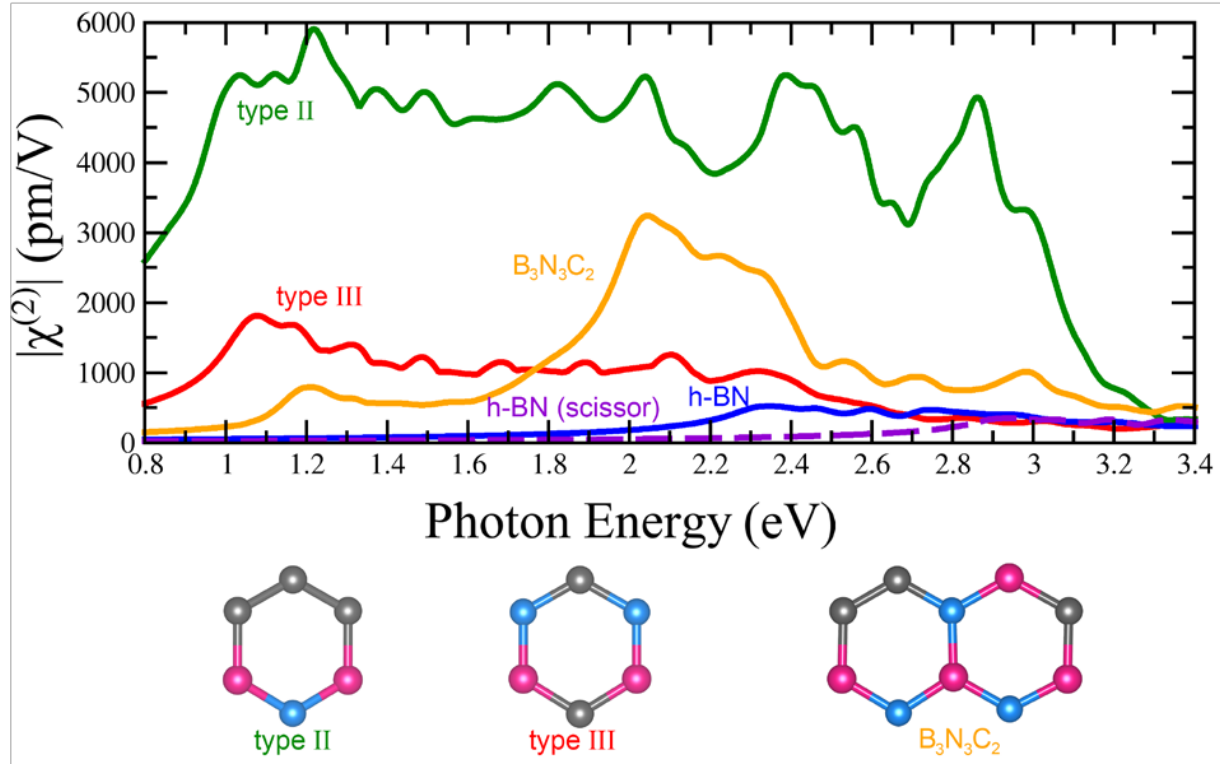

Figure 7. Upper envelopes of $|\chi^{(2)}|$ for h-BN and BNC alloys: Type II $BNC_2$, type III $BNC_2$ and $B_3N_3C_2$. H-BN scissor represents the signal when a shift in the conduction band is carried out to match the experimental band gap. The h-BN signal represents the SHG response for the DFT-LDA band gap. Models are shown where carbon atoms in grey, boron in pink, and nitrogen in blue.

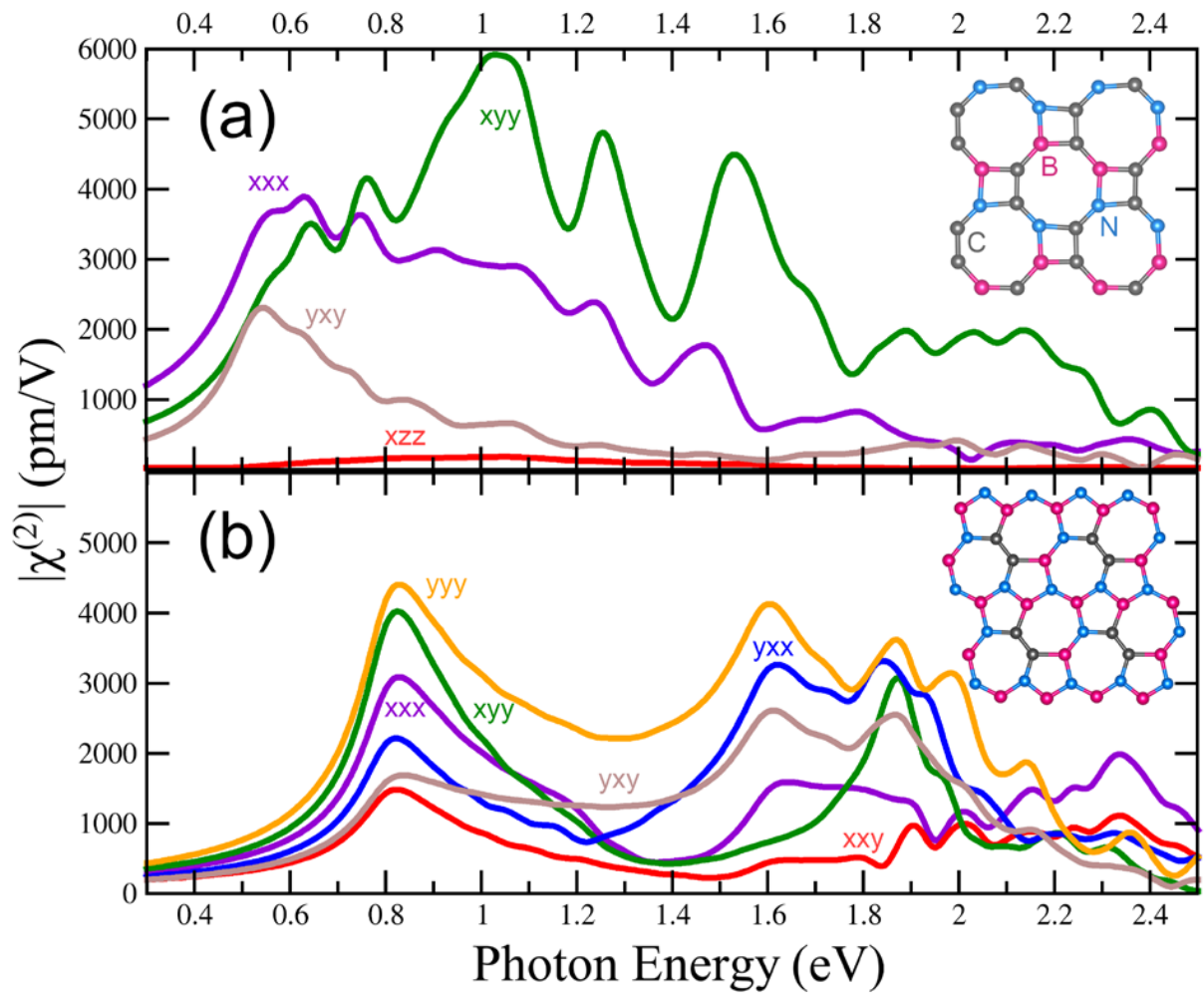

Figure 8. Second order susceptibility for BNC Haeckelites. (a) BNC-8-4 Haeckelite (squares and octagons). (b) BNC 5-7 Haeckelite (pentagons and heptagons). Carbon atoms in grey, Boron in pink, and Nitrogen in blue.

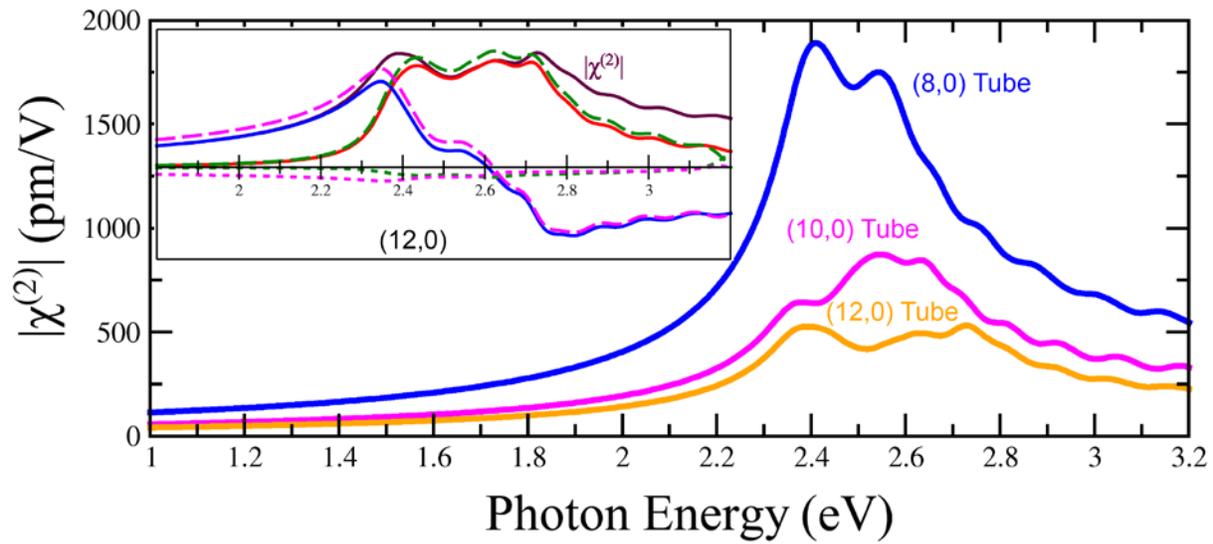

Figure 9. Comparison of $|\chi^{(2)}_{yyy}|$ for three nanotubes. inset: Comparison of the total real (solid blue) and imaginary parts (solid red) of the (12,0) BN nanotube. Real (pink) and imaginary (green) interband (dotted) and intraband (dashed) 2ω components are also shown.

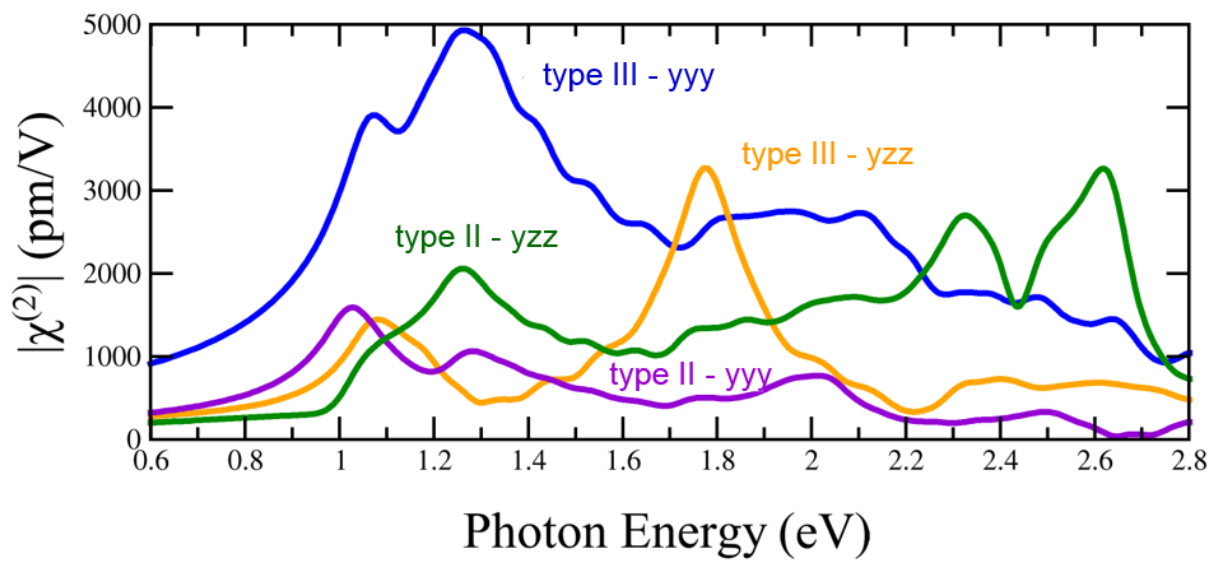

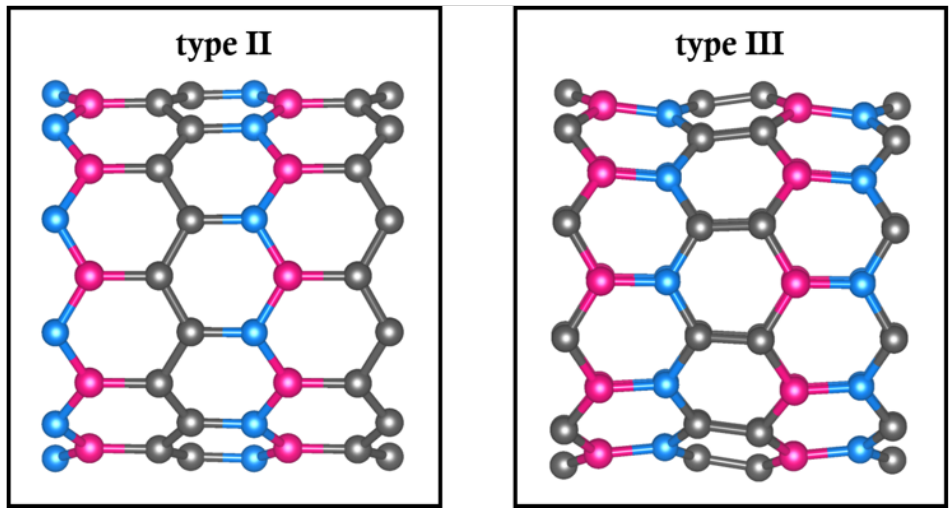

Figure 10. (a) Second order susceptibility components for two types of $BNC_2$ tubes. (b) and (c) the $BNC_2$ tubes types considered in (a): In the type II tube, B-N and C-C bonds are oriented perpendicular to the tube axis, whereas in the type III tube, they are parallel to it. Carbon atoms are grey, Boron atoms are pink, and Nitrogen atoms are blue.

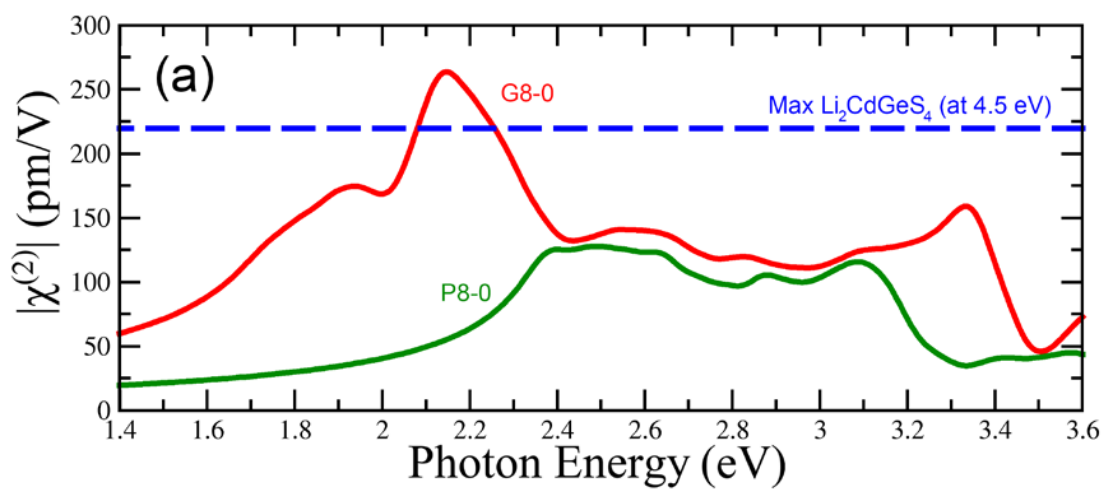
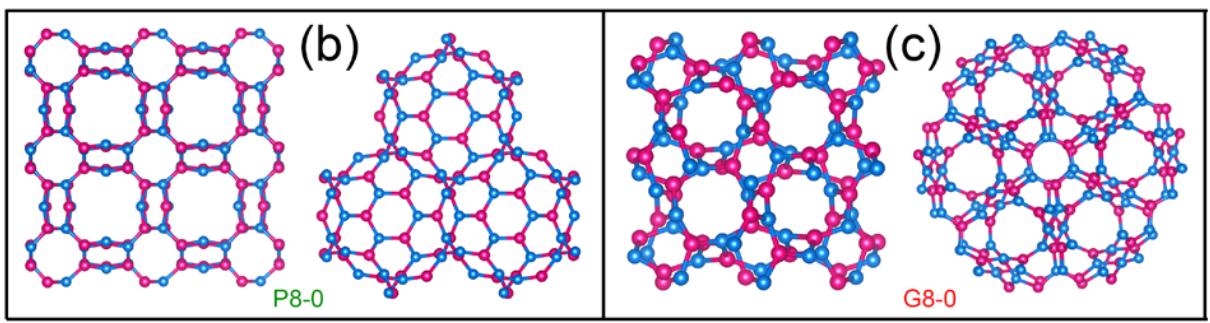

Figure 11. (a) Second order susceptibility of the G8-0 and P8-0 Schwarzites. The blue line shows the maximum value of $\chi^{(2)}$ of $Li_2CdGeS_4$ one of the highest 3-D materials. (b) and (c) Models of the BN Schwarzites: (b) P8-0. (c) G8-0 (Boron in pink and Nitrogen in blue).